\documentclass[useAMS,usenatbib]{mn2e}
\usepackage{aas_macros}
\usepackage{amsmath, amssymb}
\usepackage{graphicx}
\usepackage[caption=false,lofdepth,lotdepth]{subfig}
\usepackage{url}
\usepackage{comment}

\begin{document}

\title[Finding Lensed CCSN from Reionization]{Finding Core Collapse Supernova from the Epoch of Reionization Behind Cluster Lenses}
\author[T. Pan, A. Loeb]{Tony Pan$^1$, Abraham Loeb$^1$\\
$^1$Harvard-Smithsonian Center for Astrophysics, 60 Garden Street, Cambridge, MA 02138, USA\\
}

\pagerange{\pageref{firstpage}--\pageref{lastpage}} \pubyear{2013}
\maketitle
\label{firstpage}

\begin{abstract}

Current surveys are underway to utilize gravitational lensing by
galaxy clusters with Einstein radii $>35^{\prime\prime}$ in the search
for the highest redshift galaxies. Associated supernova from the epoch
of reionization would have their fluxes boosted above the detection
threshold, extending their duration of visibility.  We predict that
the James Webb Space Telescope (JWST) will be able to discover lensed
core-collapse supernovae at redshifts exceeding $z=7$--$8$.

\end{abstract}

\label{lastpage}

\begin{keywords}

supernovae: general -- gravitational lensing -- galaxies: clusters: general -- early Universe

\end{keywords}

\section{Introduction}

Clusters of galaxies act as gravitational lenses, focusing light-rays
from sources behind them and magnifying their images. As this effect
enables observers to probe higher redshifts than ever probed before,
surveys are being conducted with the Hubble Space Telescope (HST) to
obtain deep images of the sky through massive galaxy clusters. One
such ongoing program is the Cluster Lensing and Supernova survey
(CLASH), which is imaging 25 clusters each to a depth of 20 orbits
\citep{Postman2012}.  The 5 clusters selected for this program have
large Einstein radii of 35$^{\prime\prime}$ to 55$^{\prime\prime}$, maximizing their
potential for discovering ultra-high redshift galaxies. Indeed, three
candidate galaxies at redshifts $z\approx 9$--$10$ and another
candidate galaxy at $z\approx 11$ have already been found in the
CLASH fields \citep{Bouwens2012, Coe2013}.  Similarly, the planned HST
Frontier Fields\footnote{\url{http://www.stsci.edu/hst/campaigns/frontier-fields/}}
program will target 6 strong lensing galaxy clusters to reveal
yet higher redshifts galaxies.

The James Webb Space Telescope (JWST), the successor to HST scheduled
for launch in 2018, is likely to have analogous observational programs
with comparable integration times on a similar number of lensing
clusters.  Although the CLASH survey does aim to detect Type Ia
supernova (SN) out to redshifts of $z\sim 2.5$, the current HST
cluster observations are unlikely to detect gravitationally lensed SN
from the epoch of reionization at $z>6$. Indeed, transient science was
not identified as a science priority for the Frontier Fields program,
which will not revisit the same field twice.  The greater sensitivity
of JWST and its optimization for observations in the infrared could
potentially allow it to find lensed supernova from the cosmic dawn in
these same cluster fields.

In this {\it Letter}, we estimate the cosmic star formation rate
during the epoch of reionization by requiring that enough Pop II stars
were formed to ionize the universe.  Using model spectral time series
for Type II SN, as well as a simple isothermal sphere model for
lensing, we calculate in \S 2-5 the required magnification and
duration of detectability of such SN at $z>6$ for different JWST bands
and integration times.  Combining the above, we derive the snapshot
rate, i.e. the expected number of gravitationally lensed core collapse
SNe detected in the field-of-view of JWST around these high
magnification clusters.

\section{Star Formation \& Supernova Rate}

We infer the volumetric supernova rate $R_{SN}(z)$ as a function of
redshift by relating it to the cosmic star formation rate density
(SFRD) $\dot{\rho}_{\star}(z)$:
\begin{equation}
R_{SN}(z) = \dot{\rho}_{\star}(z) \eta_{SN} \approx \dot{\rho}_{\star}(z) \frac{\int_{M_{\rm min}}^{M_{\rm max}} \psi(M)\:dM}{0.7\int_{0.1}^{150}M\:\psi(M)\:dM},
\label{EquationRSN}
\end{equation}
where we use a Salpeter initial mass function (IMF), $\psi(M)\propto M^{-2.35}$, 
and include a factor of $0.7$ in the mass integral to
account for the shallower slope at $M\lesssim 0.5M_{\odot}$ in a
realistic IMF \citep{Fukugita1998}.  For the stellar mass range
between $M_{\rm min}=8M_{\odot}$ and $M_{\rm max}=40M_{\odot}$ appropriate for
optically-luminous core-collapse supernova, the conversion coefficient
between the star formation rate and the supernova rate is $\eta_{SN} \sim 0.0097 M_{\odot}^{-1}$.

We require that enough massive stars were formed by the end of
reionization so as to produce sufficient ionizing UV radiation to
ionize the intergalactic medium by $z_{\rm end}=6$.  This follows the
approach used in \citet{Pan2012}, albeit with different parameters to
bring our estimates closer to other inferences in literature, as
detailed below. The star formation rate during reionization peaks at
late times, when metals expelled from a prior generation of star
formation enriched the interstellar gas, so we assume that early Pop
II stars ($Z=0.02 Z_{\odot}$) with a present-day IMF dominated the
ionizing photon budget.  Using the stellar ionizing fluxes of
\citet{Schaerer2002}, we find the average number of ionizing photons
produced per baryon incorporated into a Pop II star was
$\bar{\eta}_{\gamma}=5761$.  Thus, the mass in stars per comoving
volume $\rho_{\star}(z)$ should satisfy
\begin{equation}
\rho_{\star}(z_{end}) \: \bar{\eta}_{\gamma} \: f_{esc} = C\: \rho_b,
\label{EquationRhoStarZend}
\end{equation}
where $C$ is the number of ionizing photons necessary to ionize each
baryon after accounting for recombinations, $\rho_b$ is the cosmic
baryon density, and $f_{esc}$ is the average escape fraction of
ionizing photons from their host galaxies into the intergalactic
medium.  Also, we can relate the mass in stars per volume
$\rho_{\star}(z)$ to the mass in virialized halos per volume via a
star formation efficiency $f_{\star}$:
\begin{equation}
\rho_{\star}(z) = f_{\star} \frac{\Omega_b}{\Omega_M}\int_{M_{\rm min}}^{\infty} M\:\frac{dn(z)}{dM}\:dM,
\label{EquationRhoStar}
\end{equation}
where we use the Sheth-Tormen mass function of halos for $dn/dM$
\citep{Sheth1999}, and $M_{\rm min}\sim 10^8 M_{\odot}$ is the minimum
halo mass with atomic hydrogen cooling. The cosmological parameters,
such as the matter and baryon densities $\Omega_M$, $\Omega_b$, were
taken from \citet{PlanckCollaboration2013}.
Assuming $f_{\star}$ is constant, we can calibrate $f_{\star}$ via
equations (\ref{EquationRhoStarZend}), (\ref{EquationRhoStar}), and
then evaluate $\rho_{\star}(z)$ at any redshift.  The star formation
rate is simply, $d\rho_{\star}(z)/dt$.

\begin{figure}
\centering
\includegraphics[width=1\columnwidth]{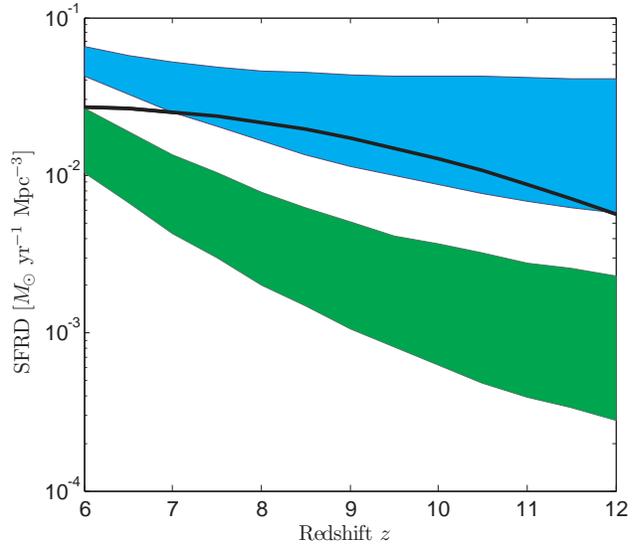}
\caption{Star formation rate density (SFRD) at high redshift.  
The black line shows our fiducial SFRD model used in later
calculations.  For comparison, the blue and green regions are taken
from \citet{Robertson2012}.  The blue region (top) spans the high and low values for parametrized
star formation histories consistent with GRB-derived star formation
rates, whereas the green region (bottom) denotes the SFRD histories
derived from UV galaxy luminosity densities observed at high redshift,
integrated down to the observation magnitude limit of $M_{AB}\approx -18$.  
Note that the latter SFRD is likely to be significantly lower
than the true cosmic SFRD, as the steep faint-end slope of lower
luminosity galaxies (possibly down to $M_{AB}\lesssim -10$) are
omitted \citep{Robertson2013,Ellis2013},
while the GRB-derived SFRD is much less flux limited and likely more
accurate.  Our SFRD parameters ($C=3$ and $f_{esc}=0.2$) were chosen
conservatively to be consistent with the low end of the GRB-derived
SFRD.}
\label{Figure_SFRD}
\end{figure}

Figure \ref{Figure_SFRD} shows our estimated SFRD, with $C=3$ and
$f_{esc}=0.2$, resulting in a ${\rm SFRD}\approx 2\times 10^{-2}
M_{\odot}$ yr$^{-1}$ Mpc$^{-3}$ (comoving) between redshifts of $z=6$
to $8$.  This corresponds to volumetric rates of approximately
$2\times 10^{-4}$ yr$^{-1}$ Mpc$^{-3}$ for core-collapse
supernova. Our simple SFRD model and the resulting SN rates linearly
scale with $C$ and $f_{esc}^{-1}$, so the JWST snapshot rates
calculated later can be easily scaled for different parameter choices
of the SFRD.

\section{Light Curves}
\label{SectionLightCurves}

\begin{figure*}
\centering
\subfloat[Subfigure 1 list of figures text][HST WFC3 1.6$\mu$m]{
\includegraphics[width=0.31\textwidth]
{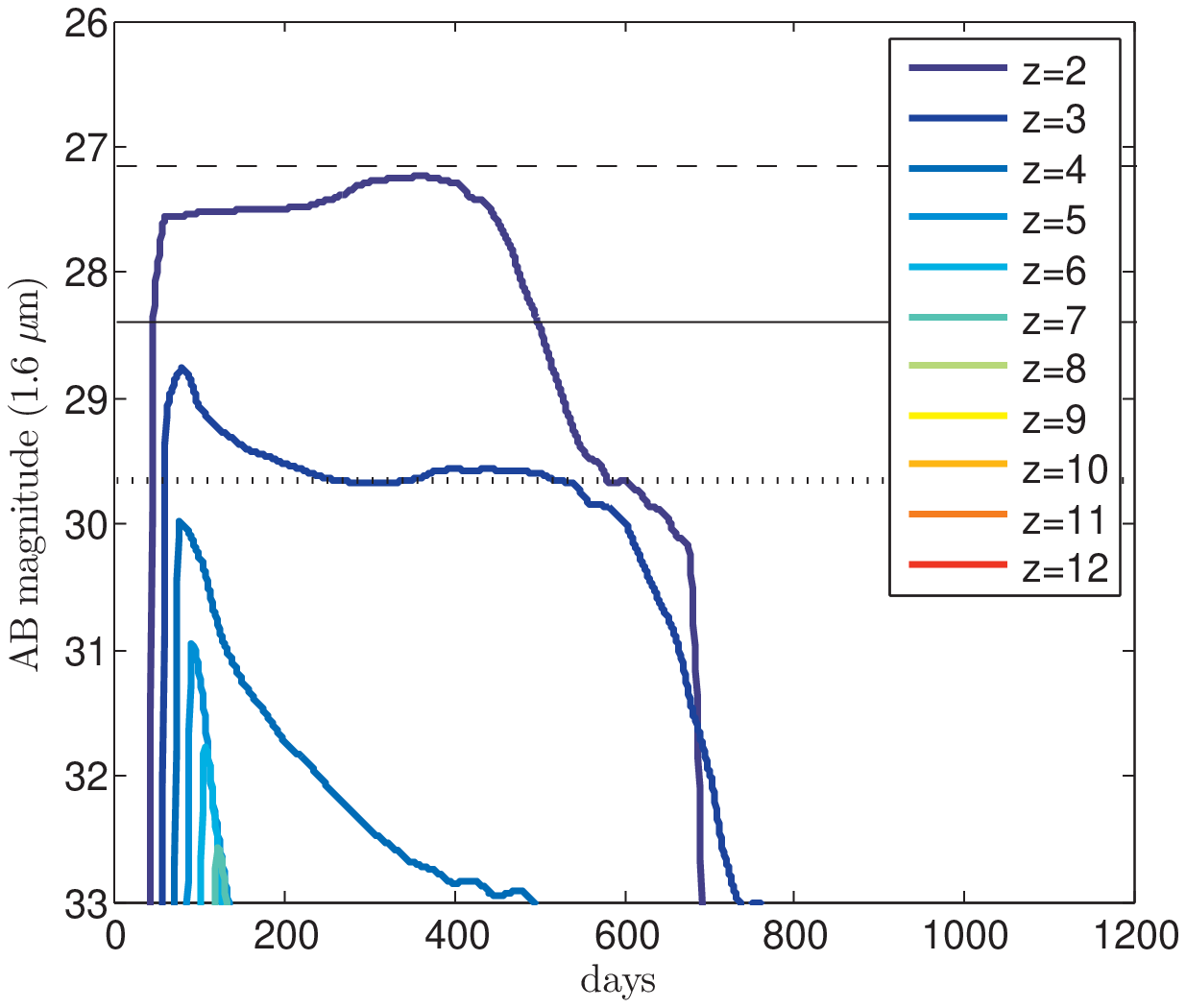}
\label{Figure_SNIIP15_LightCurve:16000}}
\quad
\subfloat[Subfigure 1 list of figures text][JWST F356W]{
\includegraphics[width=0.31\textwidth]
{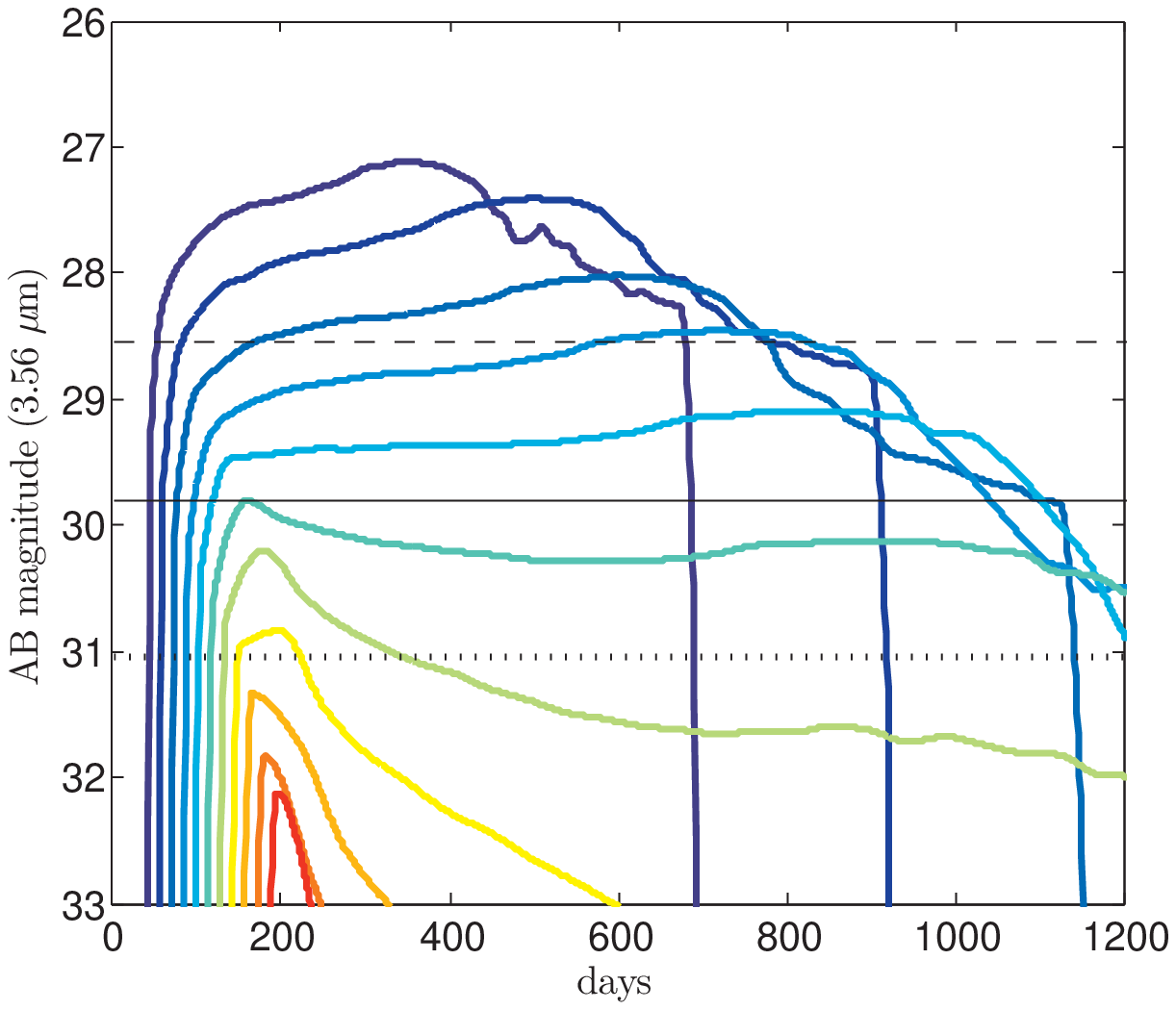}
\label{Figure_SNIIP15_LightCurve:35600}}
\quad
\subfloat[Subfigure 2 list of figures text][JWST F444W]{
\includegraphics[width=0.31\textwidth]
{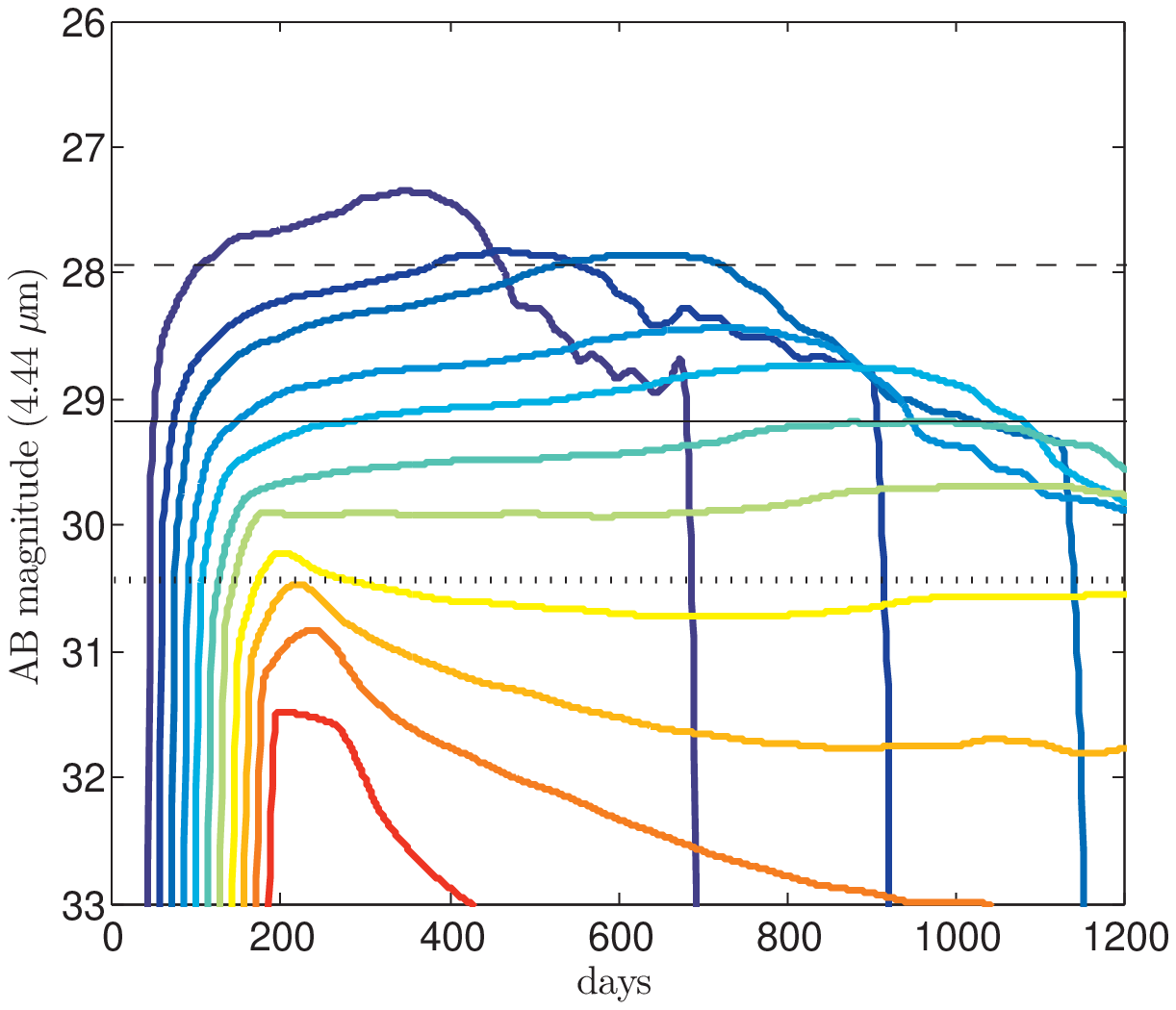}
\label{Figure_SNIIP15_LightCurve:44400}}
\caption{Observer frame light curves for a Type IIP supernova from a
$15M_{\odot}$ red giant progenitor, for the HST Wide Field Camera-3
1.6$\mu$m filter, and the JWST Near Infrared Camera (NIRCam) F356W and
F444W wideband filters at 3.56 and 4.44$\mu$m, respectively.  The
dashed, full, and dotted horizontal lines denote the AB magnitude
limits for a $10\sigma$ detection with $10^4$, $10^5$, and $10^6$s
integration times, respectively, for each filter (corresponding to
flux limits 50, 13.8, 24.5 nJy, respectively, for $10^4$s exposures).
Even a Hubble Deep Field measurement has no hope of seeing a regular
Type II SN at $z\geq 6$.  A $10^5$s exposure with JWST can detect a
$z=6$ supernova without magnification. Gravitational lensing would
extend its reach to higher redshifts and, more importantly, extend the
duration for which the supernova remains above the telescope detection
threshold.}
\label{Figure_SNIIP15_LightCurve:globfig}
\end{figure*}

We adopt the spectral time series of a Type II plateau SN from a red
giant progenitor with an initial mass $15M_{\odot}$, computed by
\citet{Kasen2009} using a code that solves the full multi-wavelength
time-dependent radiative transfer problem.  We plot the SN light
curves in the observer frame for the best possible HST and JWST
filters in Figure \ref{Figure_SNIIP15_LightCurve:globfig}.  Note that
Type II SN are diverse transients with peak luminosities that can vary
by more than an order of magnitude, and the relationship between the
progenitor mass and the brightness of the supernova is uncertain; we
adopt a single characteristic model to represent all core collapse SNe
for the sake of simplicity.  Type IIP SNe are the most common events,
and the model light curves and spectra used here agree very well with
observed SNe of average luminosities.

\begin{figure}
\centering
\includegraphics[width=1\columnwidth]
{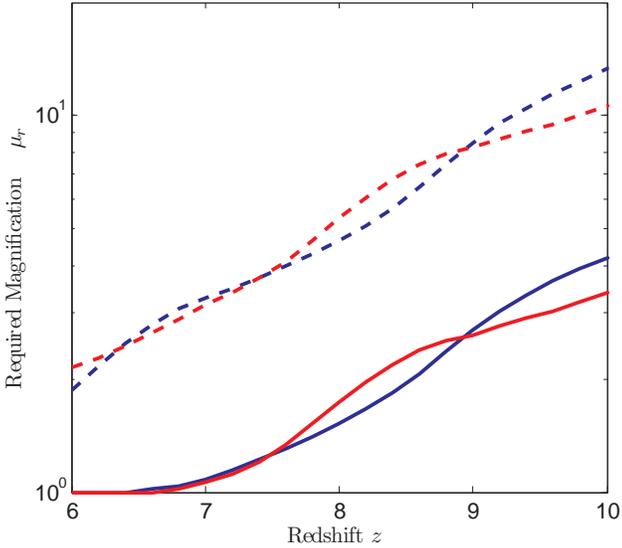}
\caption{Required magnifications $\mu_r$ for detecting Type IIP
supernovae with JWST at high redshifts.  The blue and red lines denote
the results for the F356W and F444W JWST bands, respectively, while
the dashed and solid lines correspond to integration times of $10^4$s
and $10^5$s.  The latter integration time is similar to that used in
CLASH.}
\label{Figure_Mag_vs_Zmax}
\end{figure}

We verified that HST is incapable in practice of detecting a
core-collapse SNe from the epoch of reionization.  The sensitivity of
the HST 1.6$\mu$m filter is only a factor of 2 worse than the JWST
F444W filter, but its overwhelming drawback is its waveband, which can
only probe the SN rest-frame UV flux at $z\geq 4$.  Although the JWST
F356W is more sensitive, the F444W band will be optimal for detecting
the highest redshift SN that gravitational lensing could provide.
Figure \ref{Figure_Mag_vs_Zmax} shows the magnification necessary to
detect Type II supernova at high redshifts for different integration
times.  Even with a $10^5$s exposure, a large magnification factor of
$\mu\geq 10$ will be necessary for detecting Type IIP SNe at $z>10$ with
JWST.

\section{Lensing Magnification}

For simplicity, we adopt a {\it singular isothermal sphere (SIS)}
model for the mass distribution of the lensing cluster, within which
the magnification properties are uniquely specified by the Einstein
radius $\theta_E$ \citep{Schneider1992}.  We denote the angular separations of the source
and the image from the center axis of the lens as $\beta$ and
$\theta$, respectively.  If the source lies within the Einstein radius
$\beta<\theta_E$, two images are created at locations $\theta_{\pm} =
\beta\pm\theta_E$, with magnifications $\mu_{\pm} = 1\pm
\theta_E/\beta$.  Note that $\mu_{-}$ has negative magnification, that
is, the image is flipped compared to the source.  If the source lies
outside the Einstein radius $\beta>\theta_E$, there is only one image
at $\theta=\theta_{+}$ with magnification $1<\mu_{+}<2$.
We conservatively consider only the higher-magnification image at
$\theta_{+}$, for which the source angle $\beta = {\theta_E}/({\mu-1})$.

\begin{figure}
\centering
\includegraphics[width=1\columnwidth]
{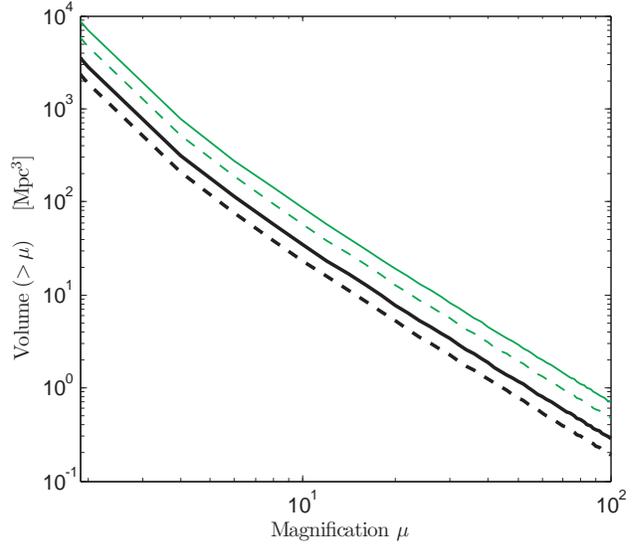}
\caption{Comoving source volume as a function of magnification $\mu$
and redshift $z$ over a redshift interval of $\Delta z =1$ for a SIS
lens.  The black and green lines denote Einstein radii of 35$^{\prime\prime}$ and
55$^{\prime\prime}$, respectively, while the solid and dashed lines
denote $z=6$ and $z=10$, respectively.  The results are in general
agreement with more realistic estimates of the search areas per
magnification factor for the magnification maps of the lensing
clusters in the CLASH survey \citep{Bouwens2012}.}
\label{Figure_SourceVolume}
\end{figure}

Then, the differential source volume (comoving) of magnified events as
a function of magnification and redshift is:
\begin{equation}
dV(z,\mu) = dA(z,\mu) \: dD_C
\label{EquationdV}
\end{equation}
where the differential comoving distance $dD_C(z)$ is
\begin{equation}
dD_C = \frac{c}{H_0} \frac{1}{E(z)} dz,
\label{EquationdD_C}
\end{equation}
with $E(z) \approx \sqrt{\Omega_M(1+z)^3+\Omega_{\Lambda}}$, and the
differential source area is
\begin{eqnarray}
\nonumber
dA(z,\mu) 	&=& \left( \: 2\pi \: D_A(z)\beta \: D_A(z)d\beta \:\right) (1+z)^2 \\
				&=& \left( 2\pi \frac{\theta_E^2}{(\mu-1)^3} d\mu \right) D_A(z)^2 (1+z)^2.
\label{EquationdA}
\end{eqnarray}
Here $D_A(z)$ is the angular diameter distance, and the extra
$(1+z)^2$ is to adjust the area to comoving units.  In Figure
\ref{Figure_SourceVolume}, we plot the source volume for a range of
Einstein radii typical of high-magnification clusters.  Given
core-collapse SN rates of $\sim 10^{-3}$ yr$^{-1}$ Mpc$^{-3}$,
capturing SN with high magnifications within source volumes $< 10^2$
Mpc$^{3}$ is unlikely.  Hence, we expect most lensed supernova
detected to have their fluxes moderately boosted with $\mu\lesssim 5$;
the benefit of lensing is to probe somewhat deeper redshifts, and to
greatly extend the duration of visibility.  Also, since high-redshift
observations are background-limited, for a target signal-to-noise
ratio, the limiting flux is proportional to $t^{-1/2}$, so even a
modest magnification of $\mu \sim 3$ can reduce the required
integration time by an order-of-magnitude.

This volume limitation of lensing also justifies our focus on
core-collapse SNe, which have the highest volumetric rates.  Although
Type Ia SNe are brighter, their volumetric rate is a factor of 4
smaller than the core collapse rate at $z\approx 7$ \citep{Pan2012},
with the difference drastically increasing with redshift due to the
long delay times needed between star formation and explosion for some
Type Ia events \citep{Maoz2012}. Pair-instability SNe from Pop III stars have volumetric rates at least
two orders of magnitude lower.

\section{Snapshot Rate}

The snapshot `rate' is the total number of events observed at a
limiting flux within a given field (not per unit time).  The
differential snapshot rate can be calculated from equations
(\ref{EquationRSN}) and (\ref{EquationdV}) via
\begin{equation}
N(z,\mu)\:dz\:d\mu= R_{SN}(z) \: t(F_{\nu},\mu,z) \: dV(z,\mu),
\label{EquationNSnapshot}
\end{equation}
where $t(F_{\nu},\mu,z)$ is the rest-frame duration over which an
event with magnification $\mu$ will be brighter than the limiting flux
$F_{\nu}$ at redshift $z$, for the observation wavelength $\nu$ under
consideration.  We find $t(F_{\nu},\mu,z)$ using our spectral
time-series for the Type IIP SN model described in \S
\ref{SectionLightCurves}.  As we care about the apparent SN rate for
observers, there is an implicit factor of $(1+z)^{-1}$ in front of the
intrinsic volumetric supernova rate $R_{SN}(z)$, but that cancels with
a $(1+z)$ factor for $t(F_{\nu},\mu,z)$ due to cosmic time dilation.

\begin{figure}
\centering
\includegraphics[width=1\columnwidth]
{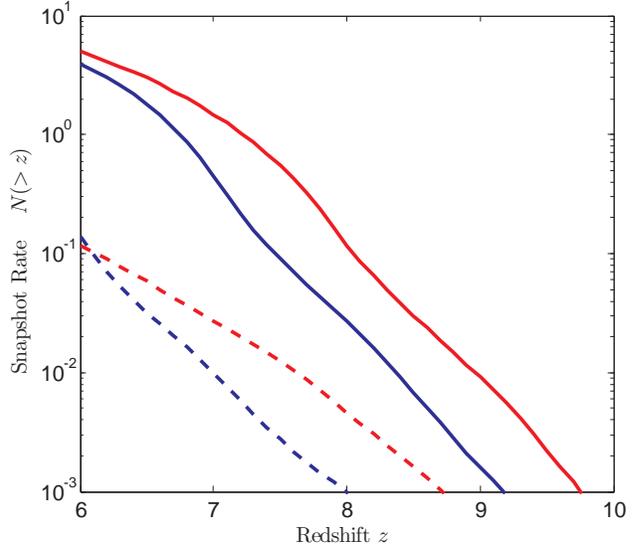}
\caption{The snapshot rate of gravitationally lensed core collapse SNe
with JWST, for a \emph{single} SIS lens with an Einstein radius
$\theta_E=35^{\prime\prime}$.  Despite the higher sensitivity of the
F356W band, the F444W band is better for finding lensed SNe at $z>6$,
as the SNe remain above the flux limit for a longer time.  Note that
5--6 high-magnification galaxy clusters with $35^{\prime\prime}\leq\theta_E\leq
55^{\prime\prime}$ are targeted in strong lensing surveys such as CLASH
and HST Frontier Fields.}
\label{Figure_Snapshot_Rate}
\end{figure}

In Figure \ref{Figure_Snapshot_Rate}, we plot the expected snapshot
rate of magnified core-collapse SN detected by JWST above target
redshifts, calculated by integrating equation
(\ref{EquationNSnapshot}) over $\mu$ and partially over $z$.  Since
NIRCam has two modules each with a 2.2$\times$2.2 arcmin$^2$
field-of-view, we limit the source area in equation (\ref{EquationdV})
to images that lie within this field-of-view.

We find that a $10^5$s JWST snapshot with the F444W filter is expected
to detect $\sim 1$ magnified core collapse SN at $z>7$ around each
cluster, and $\sim 0.1$ SNe at $z>8$.  Using $\sim 5$ clusters with
$\theta_E\geq 35^{\prime\prime}$, the prospects for detecting a few
non-superluminous SNe at high redshifts via lensing are high.  If the
other $\sim 20$ galaxy clusters in the CLASH survey with smaller
Einstein radii of $\theta_E \sim 15^{\prime\prime}$--$30^{\prime\prime}$ are also
included, the expected number of gravitationally lensed high-$z$ SNe
detected should double.

\section{Discussion}

At $z>6$, the observed duration of gravitationally-lensed
core-collapse SNe can reach $\gtrsim 1$ year, lending their detection
to a search strategy of taking images separated by $\sim 0.5-1$ year,
and looking for flux differences between consecutive snapshots.
Ideally, the cluster survey should cover most of the critical curve
area, and not just known locations of magnified images of high-$z$
galaxies, as the lensed SN may appear in currently `dark' critical
curve areas, and serve as a flag for its fainter host galaxy.  The
spectral energy distribution of Type II SNe is sufficiently different
from blackbody to allow for photometric redshift determination,
however, typing the SNe accurately may require time-consuming
spectroscopy.

Our quantitative results improve upon previous calculations of the
frequency of lensed SNe. For example, \citet{Marri2000} first explored
the effects of gravitational lensing on high-$z$ Type II SNe by
intervening cosmological mass for different cosmologies, but the
predicted detection rates were unrealistically high because of
optimistic assumptions about JWST capabilities.
\citet{Gunnarsson2003} explored the lensing by massive clusters of
distant Type Ia and Type II SNe observed at wavelengths of 0.8-1.25
microns, but found the discovery rate tapered off at $z\sim 3$.  Also,
gravitational lensing is not required per se to detect Type II SNe
from the epoch of reionization.  A moderate JWST blank-field survey
can obtain similar snapshot rates, albeit trading off the highest
redshift events for more lower redshift ones compared to a lensing
survey.  For example, \citet{Mesinger2006} found that a $10^5$s
exposure with JWST can detect 4-24 SNe per field at $z>5$, although
the assumed SFRD $\sim 0.1 M_{\odot}$ yr$^{-1}$ Mpc$^{-3}$ was an
order-of-magnitude higher than our estimates here, and the current
specifications for JWST NIRCam filter sensitivities are now $\sim 3$
times worse than the values assumed at that time.

For more luminous SNe, \citet{Whalen2012} found that core collapse SNe
from Pop III progenitors in the earliest galaxies could be visible
with the deepest JWST surveys (reaching $M_{AB}=32$) even at $z>10$,
as these SNe are bluer and almost an order-of-magnitude brighter than
the average Type II SNe considered in this paper.  \citet{Whalen2013}
also found that superluminous Type IIn SNe powered by circumstellar
interactions from Pop III stars could be visible out to $z\sim 20$.
Truly massive Pop III stars with masses $M\gtrsim 200 M_{\odot}$ can
also die as extremely bright pair-instability supernova, detectable
with JWST at $z>15$ \citep{Pan2012,Hummel2012}; indeed, the current
record for the highest-redshift supernova ever observed is likely a
pair-instability or pulsational pair-instability event at $z=3.90$
\citep{Cooke2012}.  However, the small volumetric density of Pop III
stars makes it unlikely that these events will be strongly lensed.
Finally, there is growing evidence of a prompt population of Type Ia
SNe, so their volumetric rates during the later stages of reionization
may not be negligible.  With the fiducial SFRD model in this {\it
Letter}, we estimate $\gtrsim 1$ gravitationally lensed Type Ia SNe
could be discovered at $z>7$ in the snapshots across the $\sim 5$
high-magnification clusters at any given time.

At lower redshifts, the measured core collapse SN rate is a factor of
$\sim 2$ lower than that predicted from the cosmic star formation rate
\citep{Horiuchi2011}; the most likely explanation is that some SN are
dim, whether intrinsically faint or due to dust obscuration.  This
will reduce our predicted snapshot rate.  However, we ignored the
contribution of multiple lensing images in our analysis.  Due to the
gravitational lens time delay, which could be $\sim 1-100$ years for
strong lensing around the clusters of interest \citep{Coe2013},
multiple images arriving at different times can increase the expected
snapshot detection rate of separate SN within the same field-of-view.
Although our SIS lens model can produce a maximum of only 2 magnified
images, substructure and ellipticity in actual galaxy clusters will
likely increase both the number of images and their magnifications.

\section*{ACKNOWLEDGMENTS}

We are grateful to Dan Kasen for providing the spectral time series
data for the Type IIP SN model used in this letter.  TP was supported
by the Hertz Foundation and the National Science Foundation via a
graduate research fellowship.  This work was supported in part by NSF
grant AST-0907890 and NASA grants NNX08AL43G and NNA09DB30A.

\bibliographystyle{mn2e}
\bibliography{references}

\begin{thebibliography}{22}
\expandafter\ifx\csname natexlab\endcsname\relax\def\natexlab#1{#1}\fi

\bibitem[{{Bouwens} {et~al}\mbox{.}(2012){Bouwens}, {Bradley}, {Zitrin}, {Coe},
  {Franx}, {Zheng}, {Smit}, {Host}, {Postman}, {Moustakas}, {Labbe},
  {Carrasco}, {Molino}, {Donahue}, {Kelson}, {Meneghetti}, {Jha}, {Benitez},
  {Lemze}, {Umetsu}, {Broadhurst}, {Moustakas}, {Rosati}, {Bartelmann}, {Ford},
  {Graves}, {Grillo}, {Infante}, {Jiminez-Teja}, {Jouvel}, {Lahav}, {Maoz},
  {Medezinski}, {Melchior}, {Merten}, {Nonino}, {Ogaz}, \&
  {Seitz}}]{Bouwens2012}
{Bouwens} R. {et~al.}, 2012, ArXiv e-prints

\bibitem[{{Coe} {et~al}\mbox{.}(2013){Coe}, {Zitrin}, {Carrasco}, {Shu},
  {Zheng}, {Postman}, {Bradley}, {Koekemoer}, {Bouwens}, {Broadhurst}, {Monna},
  {Host}, {Moustakas}, {Ford}, {Moustakas}, {van der Wel}, {Donahue}, {Rodney},
  {Ben{\'{\i}}tez}, {Jouvel}, {Seitz}, {Kelson}, \& {Rosati}}]{Coe2013}
{Coe} D. {et~al.}, 2013, \apj, 762, 32

\bibitem[{{Cooke} {et~al}\mbox{.}(2012){Cooke}, {Sullivan}, {Gal-Yam},
  {Barton}, {Carlberg}, {Ryan-Weber}, {Horst}, {Omori}, \&
  {D{\'{\i}}az}}]{Cooke2012}
{Cooke} J. {et~al.}, 2012, \nat, 491, 228

\bibitem[{{Ellis} {et~al}\mbox{.}(2013){Ellis}, {McLure}, {Dunlop},
  {Robertson}, {Ono}, {Schenker}, {Koekemoer}, {Bowler}, {Ouchi}, {Rogers},
  {Curtis-Lake}, {Schneider}, {Charlot}, {Stark}, {Furlanetto}, \&
  {Cirasuolo}}]{Ellis2013}
{Ellis} R.~S. {et~al.}, 2013, \apjl, 763, L7

\bibitem[{{Fukugita}, {Hogan} \& {Peebles}(1998){Fukugita}, {Hogan}, \&
  {Peebles}}]{Fukugita1998}
{Fukugita} M., {Hogan} C.~J., {Peebles} P.~J.~E., 1998, \apj, 503, 518

\bibitem[{{Gunnarsson} \& {Goobar}(2003)}]{Gunnarsson2003}
{Gunnarsson} C., {Goobar} A., 2003, \aap, 405, 859

\bibitem[{{Horiuchi} {et~al}\mbox{.}(2011){Horiuchi}, {Beacom}, {Kochanek},
  {Prieto}, {Stanek}, \& {Thompson}}]{Horiuchi2011}
{Horiuchi} S., {Beacom} J.~F., {Kochanek} C.~S., {Prieto} J.~L., {Stanek}
  K.~Z., {Thompson} T.~A., 2011, \apj, 738, 154

\bibitem[{{Hummel} {et~al}\mbox{.}(2012){Hummel}, {Pawlik},
  {Milosavljevi{\'c}}, \& {Bromm}}]{Hummel2012}
{Hummel} J.~A., {Pawlik} A.~H., {Milosavljevi{\'c}} M., {Bromm} V., 2012, \apj,
  755, 72

\bibitem[{{Kasen} \& {Woosley}(2009)}]{Kasen2009}
{Kasen} D., {Woosley} S.~E., 2009, \apj, 703, 2205

\bibitem[{{Maoz}, {Mannucci} \& {Brandt}(2012){Maoz}, {Mannucci}, \&
  {Brandt}}]{Maoz2012}
{Maoz} D., {Mannucci} F., {Brandt} T.~D., 2012, \mnras, 426, 3282

\bibitem[{{Marri}, {Ferrara} \& {Pozzetti}(2000){Marri}, {Ferrara}, \&
  {Pozzetti}}]{Marri2000}
{Marri} S., {Ferrara} A., {Pozzetti} L., 2000, \mnras, 317, 265

\bibitem[{{Mesinger}, {Johnson} \& {Haiman}(2006){Mesinger}, {Johnson}, \&
  {Haiman}}]{Mesinger2006}
{Mesinger} A., {Johnson} B.~D., {Haiman} Z., 2006, \apj, 637, 80

\bibitem[{{Pan}, {Kasen} \& {Loeb}(2012){Pan}, {Kasen}, \& {Loeb}}]{Pan2012}
{Pan} T., {Kasen} D., {Loeb} A., 2012, \mnras, 422, 2701

\bibitem[{{Planck Collaboration} {et~al}\mbox{.}(2013){Planck Collaboration},
  {Ade}, {Aghanim}, {Armitage-Caplan}, {Arnaud}, {Ashdown}, {Atrio-Barandela},
  {Aumont}, {Baccigalupi}, {Banday}, \& et~al.}]{PlanckCollaboration2013}
{Planck Collaboration} {et~al.}, 2013, ArXiv e-prints

\bibitem[{{Postman} {et~al}\mbox{.}(2012){Postman}, {Coe}, {Ben{\'{\i}}tez},
  {Bradley}, {Broadhurst}, {Donahue}, {Ford}, {Graur}, {Graves}, {Jouvel},
  {Koekemoer}, {Lemze}, {Medezinski}, {Molino}, {Moustakas}, {Ogaz}, {Riess},
  {Rodney}, {Rosati}, {Umetsu}, {Zheng}, {Zitrin}, {Bartelmann}, {Bouwens},
  {Czakon}, {Golwala}, {Host}, {Infante}, {Jha}, {Jimenez-Teja}, {Kelson},
  {Lahav}, {Lazkoz}, {Maoz}, {McCully}, {Melchior}, {Meneghetti}, {Merten},
  {Moustakas}, {Nonino}, {Patel}, {Reg{\"o}s}, {Sayers}, {Seitz}, \& {Van der
  Wel}}]{Postman2012}
{Postman} M. {et~al.}, 2012, \apjs, 199, 25

\bibitem[{{Robertson} \& {Ellis}(2012)}]{Robertson2012}
{Robertson} B.~E., {Ellis} R.~S., 2012, \apj, 744, 95

\bibitem[{{Robertson} {et~al}\mbox{.}(2013){Robertson}, {Furlanetto},
  {Schneider}, {Charlot}, {Ellis}, {Stark}, {McLure}, {Dunlop}, {Koekemoer},
  {Schenker}, {Ouchi}, {Ono}, {Curtis-Lake}, {Rogers}, {Bowler}, \&
  {Cirasuolo}}]{Robertson2013}
{Robertson} B.~E. {et~al.}, 2013, ArXiv e-prints

\bibitem[{{Schaerer}(2002)}]{Schaerer2002}
{Schaerer} D., 2002, \aap, 382, 28

\bibitem[{{Schneider}, {Ehlers} \& {Falco}(1992){Schneider}, {Ehlers}, \&
  {Falco}}]{Schneider1992}
{Schneider} P., {Ehlers} J., {Falco} E.~E., 1992, {Gravitational Lenses}

\bibitem[{{Sheth} \& {Tormen}(1999)}]{Sheth1999}
{Sheth} R.~K., {Tormen} G., 1999, \mnras, 308, 119

\bibitem[{{Whalen} {et~al}\mbox{.}(2013){Whalen}, {Even}, {Lovekin}, {Fryer},
  {Stiavelli}, {Roming}, {Cooke}, {Pritchard}, {Holz}, \&
  {Knight}}]{Whalen2013}
{Whalen} D.~J. {et~al.}, 2013, ArXiv e-prints

\bibitem[{{Whalen} {et~al}\mbox{.}(2012){Whalen}, {Joggerst}, {Fryer},
  {Stiavelli}, {Heger}, \& {Holz}}]{Whalen2012}
{Whalen} D.~J., {Joggerst} C.~C., {Fryer} C.~L., {Stiavelli} M., {Heger} A.,
  {Holz} D.~E., 2012, ArXiv e-prints

\end{thebibliography}

\end{document}